\documentclass[aps,pre,preprint]{revtex4-1}

\usepackage{graphicx}

\begin{document}

\title{Ergodicity in a two-dimensional self gravitating many body system}
\author{C.~H.~Silvestre}
\affiliation{Instituto de F\'\i{}sica -- Universidade de
Bras\'\i{}lia\\ CP: 04455, 70919-970 - Bras\'\i{}lia, Brazil}
\author{T.~M.~Rocha Filho}
\affiliation{Instituto de F\'\i{}sica and International Center for Condensed Matter Physics\\ Universidade de
Bras\'\i{}lia, CP: 04455, 70919-970 - Bras\'\i{}lia, Brazil}

\begin{abstract}
We study the ergodic properties of a two-dimensional self-gravitating system using molecular dynamics simulations. We apply three different
tests for ergodicity: a direct method comparing the time average of a particle momentum and position to the respective ensemble average,
sojourn times statistics and the dynamical functional method. For comparison purposes they are also applied to a short-range interacting system
and to the Hamiltonian mean-field model. Our results show that a two-dimensional self-gravitating system takes a very long time to establish ergodicity.
If a Kac factor is used in the potential energy, such that the total energy is extensive, then this time is independent
of particle number, and diverges with $\sqrt{N}$ without a Kac factor.
\end{abstract}


\maketitle

\section{Introduction}

$N$-body systems with long-range interactions have some peculiarities with respect to systems with short-range interactions, having
drawn much attention at least along the last two decades~\cite{newbook,proc1,proc2,proc3,physrep}, and for a longer time if one considers
self-gravitating systems and charged plasmas. Starting from a non-equilibrium configuration, a short-range interacting system
evolves to thermodynamic equilibrium in a relatively small relaxation time, while a long-range interacting system evolves over different stages characterized
by time-scales differing by orders of magnitude, and taking a very long time to reach thermodynamic equilibrium for a finite number of particles.
The initial stage of evolution is a violent relaxation
into a Quasi-Stationary State (QSS) in a time roughly independent on the number of particles~\cite{lyndenbell}. The relaxation time
to thermodynamic equilibrium diverges in the $N\rightarrow\infty$ limit (Vlasov limit) and gets trapped in a non-equilibrium non-Gaussian stationary state.

A pair interaction potential is long-ranged if
it decays at long distances as $r^{-\gamma}$ with $\gamma<d$, with $d$ the spatial dimension and $r$ the inter-particle distance.
This implies that all particles, no matter how far, contribute to the total energy. Consequently
the system is non-additive, violating the fundamental axiom of thermodynamics: non-additivity of entropy.
It is worth remembering that this is not in contradiction with the second law of Thermodynamics, which is always valid~\cite{latella}.
Another common consequence is non-ergodicity, most extensively studied
for the Hamiltonian mean field model~\cite{hmforig,mukamel,nosepl,bouchet,chavanis,levin,pakter,ribeiro},
but also for one and three-dimensional self gravitating systems~\cite{yawn,yawn2,ping,teles0,teles0b}.
Quite curiously, and up to the authors knowledge, no previous detailed study was devoted to the ergodicity of the two-dimensional case,
at least not in the sense discussed here.
In the Vlasov limit, the dynamics of the system is exactly described by the Vlasov equation, which is essentially the Liouville equation
for the one-particle distribution function evolving in the mean-field due to all other particles, and given by~\cite{braun,nosval}:
\begin{equation}
\frac{d}{dt}f({\bf p},{\bf r};t)=\left(\frac{\partial}{\partial t}+\frac{\bf p}{m}\cdot\frac{\partial}{\partial\bf r}
+{\bf F}({\bf r};t)\cdot\frac{\partial}{\partial\bf p}\right)=0,
\label{vlasoveq}
\end{equation}
where $f({\bf p},{\bf r};t)$ is the one-particle distribution function, ${\bf p}$ and ${\bf r}$ the momentum and position vectors in $d$ spatial dimensions,
respectively, and
\begin{equation}
{\bf F}({\bf r};t)=-\frac{\partial}{\partial\bf r}\int {\rm d}{\bf r}^\prime\: V({\bf r}-{\bf r}^\prime),
\label{meanfforce}
\end{equation}
is the mean field force at position ${\bf r}$ and time $t$ with $V({\bf r}-{\bf r}^\prime)$ the inter-particle potential.
For finite $N$ the right-hand side of Eq.~(\ref{vlasoveq}) does no longer vanish and is given by the pertinent collisional contributions
(granularity effects)~\cite{balescu}.
%%%%%%%%%%%%%%%%%%%%%%%%%%%%%%%%%%%%%%%%%%%%%%%%%%%%%%%%%%%%%%%%%%%%%%%%%%%%%%%%%%%%%%%%%%%%%%%%%%%%%%%%%%%
For a long-range interacting system the thermodynamic limit is ill defined~\cite{ruelle}, and the Vlasov limit can be taken consistently
by introducing a Kac factor $1/N$ in the interaction potential, such that the energy becomes extensive, but still non-additive,
and that the $N\rightarrow\infty$ limit converges~\cite{kac}.
%%%%%%%%%%%%%%%%%%%%%%%%%%%%%%%%%%%%%%%%%%%%%%%%%%%%%%%%%%%%%%%%%%%%%%%%%%%%%%%%%%%%%%%%%%%%%%%%%%%%%%%%%%%

The study of ergodicity was pioneered by Boltzmann in his works on the foundations on Statistical Mechanics~\cite{Max}, and latter extended
by Birkhoff~\cite{birkhoff} and Khinchin~\cite{Stam}. The ergodic hypothesis states that
time average of a dynamical functions $b(x)$ equals the ensemble average:
\begin{equation}
\lim_{t_f\rightarrow\infty}\frac{1}{t_f-t_0}\int_{t_0}^{t_f} {\rm d}t\: b(x_t)\equiv\overline{b}(x)=\langle b(x)\rangle\equiv \int {\rm d}\mu_0\: b(x),
\end{equation}
where $x$ denotes a point in the system state space (or phase space for a Hamiltonian classical system),
$\overline{b}(x)$ is the time average of $b(x)$, $\langle b(x)\rangle$ its ensemble average and $d\mu_0$ a statistical measure.
Proving ergodicity rigorously is a difficult task and has been accomplished only for a few cases. Most studies rely on different
methods such as determining the existence of gaps in phase-space~\cite{mukamel,bouchet}, direct comparison of time and ensemble averages
for the momentum variable~\cite{nosepl}, sojourn time statistics for cells in phase-space~\cite{weba,venegeroles,prezol},
testing for equipartition of energy~\cite{yawn,yawn2,tsuchiya} and the dynamical functional approach~\cite{janicki,cambanis,magdziarz}. The latter
can also be used to determined weather the system is mixing, which is a stronger property than ergodicity.
The dynamics of a system is mixing in its phase-space $S$ if for an invariant measure $\mu$, ${\bf T}$ a map preserving $\mu$ and $s_1,s_2\subset S$,
we have that~\cite{ott}:
\begin{equation}
\frac{\mu(s_1)}{\mu(S)}=\lim_{k\rightarrow\infty}\frac{\mu(s_2\cap{\bf T}^k s_1)}{\mu(s_2)}.
\label{misdef}
\end{equation}
Non-ergodicity can also be classified as strong if some regions of
phase-space are non-accessible to the system and weak if all regions are accessible but not equally visited.

Ergodicity can be studied in two complementary ways. The first possibility is to consider the evolution in phase space of the $N$-particle system,
with time averages computed over the trajectory of the point representing the state of the whole system. Ensemble averages are
then obtained by considering all points in the energy hypersurface as equally probable, i.~e.\ for the microcanonical equilibrium.
This is equivalent to state that the total time spent in a phase-space region is asymptotically proportional
to its measure.
The other possibility is to consider single particle histories as realizations of a stochastic process.
In this case time averages are taken over one particle history, while ensemble averages are taken
over the set of all $N$ particles. This approach is particularly suited for a direct experimental verification even for short-range interacting systems~\cite{feil}.
In this case ergodicity, i.~e.\ the coincidence of the time and ensemble averages is not equivalent to thermodynamic equilibrium
as will become evident below. We consider the latter case in the present work, with $S$ in Eq.~(\ref{misdef}) the one-particle phase-space,
for a system composed by a single particle evolving in a stationary potential, either in the thermodynamic equilibrium or in a (quasi-) stationary state.

In this paper we first briefly study a gas of elastic hard-discs as an example of short-range interacting system, and then
revisit the ergodic properties of the Hamiltonian Mean Field (HMF) model extending the results of Ref.~\cite{nosepl} to include non-homogeneous
quasi-stationary states. Both system are used as comparison standards for our main interest here, a two-dimensional self-gravitating $N$-body system.
The HMF model is formed by $N$ particles on a circle with Hamiltonian~\cite{hmforig}:
\begin{equation}
H=\sum_{k=1}^N\frac{p_k^2}{2}+\frac{1}{2N}\sum_{k,l=1}^N\left[1-\cos\left(\theta_k-\theta_l\right)\right],
\label{hamhmf}
\end{equation}
where $\theta_k$ is the position angle on particle $k$ and $p_k$ its conjugate momentum.
It is a solvable model at equilibrium~\cite{hmforig,metodo} at its dynamics can be simulated with a smaller
computational cost if compared to other long-range interacting systems~\cite{eugpumd} and widely studied in the literature
(see~\cite{newbook,proc1,proc2,proc3} and references therein).
A two-dimension self-gravitating system with $N$ identical particles, as described by the Hamiltonian:
\begin{equation}
H=\sum_{k=1}^N\frac{p_k^2}{2}+\frac{1}{2N}\sum_{k,l=1}^N\ln\left(\left|{\bf r}_k-{\bf r}_l\right|+\epsilon\right),
\label{ham2dgrav}
\end{equation}
where the logarithmic potential is the solution of the Poisson equation in two dimensions~\cite{levipprep}, all masses are set to unity and
$\epsilon$ is a softening small parameter, commonly used in simulations of self-gravitating systems to avoid the divergence in the potential
at zero distance~\cite{aarseth}. Note that we use a Kac factor $1/N$ in the potential.

The potential in Eq.~(\ref{ham2dgrav}) is not upper bound and is therefore confining, consequently avoiding the difficulty
of particle evaporation in three-dimensional gravity. This model was used for instance in the study of anomalous diffusion in a collapsing phase~\cite{antoni2},
the determination of thermodynamic equilibrium properties~\cite{aly}, collisional relaxation~\cite{marcos,nosval},
violent relaxation~\cite{levipprep,teles} and cooling in self-gravitating accretion discs~\cite{faghei}.

The structure of the paper is as follows: In section~\ref{te} we succinctly present the methods for testing ergodicity used in the present work,
and apply then to a two-dimensional hard-disc gas in Section~\ref{sris}, and to the HMF model in Section~\ref{hmfrev}.
The ergodic properties of a two-dimensional self-gravitating system are discussed in Section~\ref{tdg}.
We close the paper with some concluding remarks in Section~\ref{cr}.

\section{Testing ergodicity}
\label{te}

\subsection{Direct method}
\label{dirmeth}

We consider the time evolution of a single particle to compute time averages while ensemble averages are obtained by taking the average over
the $N$ particles in the whole system. For an ergodic system both quantities coincide. Let us then consider
the momentum $p_k(t)$ of particle $k$ and its position $x_k(t)$ at time $t$ and write their time averages as:
\begin{equation}
\overline{p}_k(t)=\frac{1}{n}\sum_{j=0}^n p_k(j\Delta t),
\label{drt1}
\end{equation}
and
\begin{equation}
\overline{x}_k(t)=\frac{1}{n}\sum_{j=0}^n x_k(j\Delta t),
\label{drt1b}
\end{equation}
respectively, where $\Delta t$ is a constant time interval, that we take as being the numeric integration time step and $m$ is the total number of such intervals.
The time averages are computed up to a given time $t$ denoted by the argument of $\overline{p}_k(t)$ and $\overline{x}_k(t)$.
On the other hand, ensemble averages at time $t$ are given by
\begin{equation}
\langle p(t)\rangle=\frac{1}{N}\sum_{i=0}^N p_i(t),
\label{drt2}
\end{equation}
and
\begin{equation}
\langle x(t)\rangle=\frac{1}{N}\sum_{i=0}^N x_i(t).
\label{drt2b}
\end{equation}
For an ergodic system $\overline{p}_k$ and $\overline{x}_k$ are the same for all particles, and therefore the standard deviations:
\begin{equation}
\sigma_{\overline{p}}(t)=\sqrt{\frac{1}{N}\sum_{k=1}^{N} \overline{p}_k(t)^2-\langle \overline{p}(t)\rangle^2},
\label{drt3}
\end{equation}
and
\begin{equation}
\sigma_{\overline{x}}(t)=\sqrt{\frac{1}{N}\sum_{k=1}^{N} \overline{x}_k(t)^2-\langle \overline{x}(t)\rangle^2},
\label{drt3b}
\end{equation}
must vanish asymptotically with time. In Eqs.~(\ref{drt3}) and~(\ref{drt3b}) we used $\langle \overline{p}(t)\rangle\equiv(1/N)\sum_k\overline{p}_k(t)$
and $\langle \overline{x}(t)\rangle\equiv(1/N)\sum_k\overline{x}_k(t)$.

The direct method then consists in verifying numerically
that $\sigma_{\overline{p}}$ and $\sigma_{\overline{x}}(t)$ do indeed vanish (or approach zero) for some sufficiently long time $t$~\cite{nosepl}.
For more than one spatial dimension this must be true for each momentum and position components.

It is interesting to note the connection of ergodicity breaking and anomalous diffusion in long-range interacting systems.
In the limit $\Delta t\rightarrow 0$ Eq.~(\ref{drt1}) is rewritten as:
\begin{equation}
\overline{p}_k=\frac{1}{t}\int_0^t {\rm d}t\:p_k(t)=\frac{m}{t}\int _{x_k(0)}^{x_k(t)}{\rm d}x_k=\frac{m}{t}\left[x_k(t)-x_k(0)\right].
\label{drt1c}
\end{equation}
Plugging Eq.~(\ref{drt1c}) into Eq.~(\ref{drt3}) and assuming $\langle p(t)\rangle=0$ we obtain:
\begin{equation}
\sigma_{\overline{p}}(t)=\frac{m}{t}\sqrt{\frac{1}{N}\sum_{k=1}^{N} \left[x_k(t)-x_k(0)\right]^2}.
\label{drt3c}
\end{equation}
Anomalous diffusion is ubiquitous in systems with long-range interactions, with the standard deviation of the distance traveled by a particle
satisfying:
\begin{equation}
\frac{1}{N}\sum_{k=1}^{N} \left[x_k(t)-x_k(0)\right]^2=C\hspace{1pt}t^\mu,
\label{drt3d}
\end{equation}
with $C$ and $\mu$ constants, and the case $\mu=1$ corresponding to normal diffusion. From Eqs.~(\ref{drt3}), (\ref{drt1c}) and~(\ref{drt3c}) we have
\begin{equation}
\sigma_{\overline{p}}(t)=mC\hspace{1pt}t^{\mu/2-1}.
\label{drt1d}
\end{equation}
As a consequence the system is always ergodic in the momentum variable whenever $\mu<2$. Nevertheless the observation time required
to reach ergodicity can be very long for $\mu$ smaller than but close to $2$. This time also diverges if transport is ballistic or if
$C\rightarrow\infty$ for $N\rightarrow\infty$. A similar approach is not possible for the position variable
as in this case there is no relation analogous to Eq.~(\ref{drt1c}). Anomalous diffusion in the HMF model was thoroughly studied
in the literature~\cite{antoni2,latora,yamaguchi,sire,bouchet2,moyano,antoniazzi}, although some points still deserve more investigation.
For self-gravitating systems see Ref.~\cite{fouvry} and references therein.

\subsection{Sojourn time statistics}
\label{sojtimessec}

This approach was introduced by Rebenshtok and Barkai for a Continuous Time Random Walk (CTRW)~\cite{weba}. Consider a system of
discrete states labeled by an integer index, each one visited intermittently for a given time. The time spent in state $k$ during the $j$-th visitation
is denoted by $t_{k,j} ^{(s)}$ and called a sojourn time. The sum over all sojourn times for a given state yields its residence time:
\begin{equation}
t_k = \sum _j t_{k,j} ^{(s)}. 
\end{equation}
The probability density function for a weak non-ergodic system for the time average $\overline{\cal O}$ of an observable ${\cal O}$ can be written as~\cite{weba}:
\begin{equation}
\label{dist}
f ^{(\alpha)} (\overline{\cal O})= \frac{1}{\pi} \lim_{\epsilon\rightarrow 0}\:{\rm Im}\left[\frac{\sum_{k=1}^{L} p_k^{eq}(\overline{\cal O}- \overline{\cal O}_k
+i\epsilon)^{\alpha/2-1}}{\sum_{k=1}^{L} p_k^{eq}(\overline{\cal O}- \overline{\cal O}_k + i\epsilon)^{\alpha/2}}\right],
\end{equation} 
where $\alpha$ is a constant in the interval $(0,2]$, ${\cal O}_k$ is the value of the observable in state $k$, $p_k^{eq}$
is the probability for the system to be in the state $k$ and $L$ the number of discrete states.
The time average of the observable ${\cal O}$ is thus:
\begin{equation}
\overline{\cal O}=\frac{1}{t_{tot}}\sum_{k=1}^L t_k {\cal O}_k,
\end{equation}
with $t_{tot}= \sum_{i=k}^L t_k$. If the statistics of the sojourn times $t_{k,j} ^{(s)}$ have a power law tail,
then the system is non-ergodic. Otherwise the system is ergodic and we have that $\alpha=2$ in Eq.~(\ref{dist}) according to the classical central limit theorem.
This implies $f ^{(\alpha)} (\overline{\cal O})\rightarrow\delta(\overline{\cal O}-\langle{\cal O}\rangle)$ and consequently
\begin{equation}
\label{gaus}
f ^{2} (\overline{O})= \delta (\overline{O} - \langle \overline{O} \rangle).
\end{equation} 
Usually for a finite system the time a particle spends in a given cell cannot be arbitrarily large, and
the distribution of sojourn times must be truncated at some value of time. For the Hamiltonian mean field model
the statistics of sojourn times exhibits a truncated algebraic tail with a truncation time diverging with $N$,
implying that the system is non-ergodic only in the limit $N\rightarrow\infty$~\cite{prezol}.

\subsection{Dynamical functional method}

For a stationary
infinitely divisible processes $Y(n)$, with $n$ integer, the dynamical functional is given by the Fourier transform with unit wave number
of the process $Y(n)-Y(0)$~\cite{janicki,cambanis,magdziarz}:
\begin{equation}
D(n)=\left\langle e^{i[Y(n)-Y(0)]}\right\rangle,
\end{equation}
where $\langle\cdots\rangle$ denotes the ensemble average, i.~e.\ an average over many realizations of the stochastic process $Y(n)$.
The system is ergodic if and only if~\cite{magdziarz}:
\begin{equation}
\lim_{n\rightarrow \infty}\frac{1}{n}\sum_{k=1}^{n-1}D(k)={|\langle e^{iY(0)}\rangle |}^2,
\label{ergodiccond}
\end{equation}
and mixing if and only if:
\begin{equation}
\lim_{n\rightarrow \infty} D(n)={|\langle e^{iY(0)}\rangle |}^2.
\label{mixcond}
\end{equation}
We identify $Y(n)$ to the momentum or position variables (or any of their components) at time $n\Delta t$ for a fixed time interval $\Delta t$,
such that
\begin{equation}
\left\langle e^{i\left[Y(n)-Y(0)\right]}\right\rangle=\frac{1}{N}\sum_{k=1}^N e^{i\left[p_k(n\Delta t)-p_k(0)\right]},
\label{nova1}
\end{equation}
and similarly for the position variable.

It is useful to define the new functions:
\begin{eqnarray}
 & & E(n)\equiv D(n) - {|\langle e^{iY(0)}\rangle |}^2,
\nonumber\\
 & & Q(n)\equiv\frac{1}{n}\sum_{k=1}^{n-1}E(k).
\label{efen}
\end{eqnarray}
Then Eqs.~(\ref{ergodiccond}) and~(\ref{mixcond}) are rewritten, respectively, as:
\begin{equation}
\lim_{n\rightarrow \infty}Q(n)=0,
\label{ergodiccond2}
\end{equation}
and
\begin{equation}
\lim_{n\rightarrow \infty} E(n)=0.
\label{mixcond2}
\end{equation}
Testing for mixing is of special interest as many theories for violent relaxation presuppose a good mixing in the initial evolution~\cite{lyndenbell}
(see also~\cite{corehalomeu} and references therein).

We define now the time for ergodicity as the time scale such that the difference in time and ensemble averages are negligible, and can be estimated
using the direct method, which simply implements the definition of ergodicity. The truncation of sojourn times and the value of $n$ in Eqs.~(\ref{ergodiccond2})
and~(\ref{mixcond2}) at which $E(n)$ and $Q(n)$ are close to zero, although somehow related to the time of ergodicity, are not a priori the same.

\section{Short-range interacting system - Hard-discs in 2D}
\label{sris}

\begin{figure}[ht]
\begin{center}
\includegraphics[width=16cm]{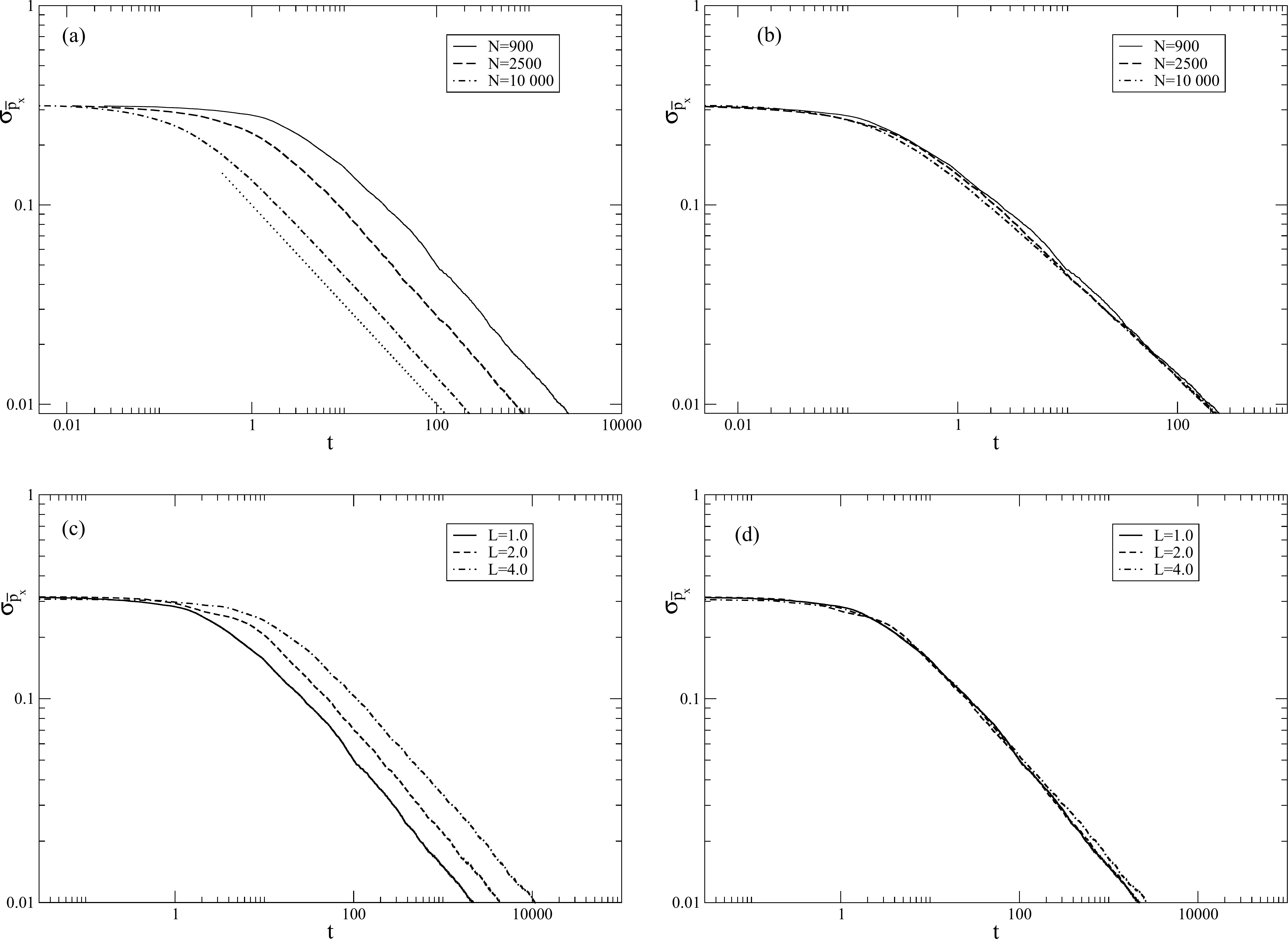}
\end{center}
\caption{a) Standard deviation $\sigma_{\overline{p}_x}$ in Eq.~(\ref{drt3}) of the time average of the $x$-component of the momentum of
the computed over all particles for a gas of hard-discs
for $N=900$, $N=2500$ and $N=10\,000$, with disc diameter $d=0.002$ and inverse temperature $\beta=10.0$. The dotted line is the curve proportional to $1/\sqrt{t}$.
b) Same as (a) but with time rescaled as $t\rightarrow t\times N/10\,000$. The good data collapse indicates that $\sigma_{\overline{p}_x}(t)\sim N^{-1}$.
c) Standard deviation $\sigma_{\overline{p}_x}$ for $N=900$ but different values for the length of the side $L$ of the square box with $d=0.002$.
d) Same as (c) but with time rescaled as $t\rightarrow t/L$. The good data collapse indicates that $\sigma_{\overline{p}_x}(t)\sim L$.}
\label{sigshort}
\end{figure}

As an illustration of the methods described in the previous section and for comparison purposes, let us
consider a two-dimensional system composed of identical hard-discs of unit mass and diameter $d$ with instantaneous (contact) elastic collisions.
Simulations are performed using an event-driven algorithm~\cite{allen}.
The system is prepared in a homogeneous distribution in a square box with sides of length $L=1$ and a Gaussian (equilibrium) initial distribution of velocities
with inverse temperature $\beta$. For economy of space we only perform for this model ergodicity tests for the dynamics of the $x$-component of the momentum variable,
as the position variables have a similar behavior. To apply the direct method we compute the time average of the momentum
of each particle and then compute the standard deviation $\sigma_{\overline{p}_x}(t)$ of these averages over all $N$ particles, as given by
Eq.~(\ref{drt3}). Results for $N=900$, $N=2500$ and $N=10\,000$, disc diameter $d=0.002$ and $\beta=10.0$ are shown in Fig.~\ref{sigshort}a.
The dependence of the tail of $\sigma_{\overline{p}_x}(t)$ on time and $N$
can be determined by considering that increments momentum due to collisions are uncorrelated, implying an asymptotic time dependence
in $t^{-1/2}$ as a consequence of the classical central limit theorem, and that the collisions frequency is proportional to $N$, i.~e\
that $\sigma_{\overline{p}_x}(t)\sim 1/N\sqrt{t}$ as is visible in Fig.~\ref{sigshort}b. The thermodynamic limit
$N\rightarrow\infty$, $L^2\rightarrow\infty$ with $N/L^2={\rm constant}$, can be inferred from Figs.~\ref{sigshort}c and~\ref{sigshort}d,
where some simulations with $N=900$ and some values of $L$ are shown. It becomes clear that $\sigma_{\overline{p}_x}(t)\sim L$. From the former results
we conclude that:
\begin{equation}
\lim_{N\rightarrow\infty}\sigma_{\overline{p}_x}(t)=0,\hspace{5mm}\frac{N}{L}={\rm constant}.
\label{thlimhard}
\end{equation}
Therefore, the time for ergodicity vanishes in the thermodynamic limit.

The average thermal velocity can be defined at equilibrium by $\overline{v}\equiv\sqrt{\langle v^2\rangle}=1/\sqrt{\beta}$,and 
two different time-scales can be defined from it: the average time between two collisions for a given particle $\tau_{col}=l/\overline{v}$,
with $l$ the mean free path which is roughly proportional to $\sqrt{N}$, and the crossing time, i.~e.\ the time for a particle with average thermal
velocity to cross the box $\tau_{cr}=1/\overline{v}$. For the state considered in Fig. \ref{sigshort} we have $\overline{v}\approx0.32$.
The time required for $\sigma_{\overline{p}_x}(t)$ to decrease by two orders of magnitude is roughly
two order of magnitudes bigger than the crossing time $\tau_{cr}=\sqrt{10}$.

\begin{figure}[ht]
\begin{center}
\includegraphics[width=16cm]{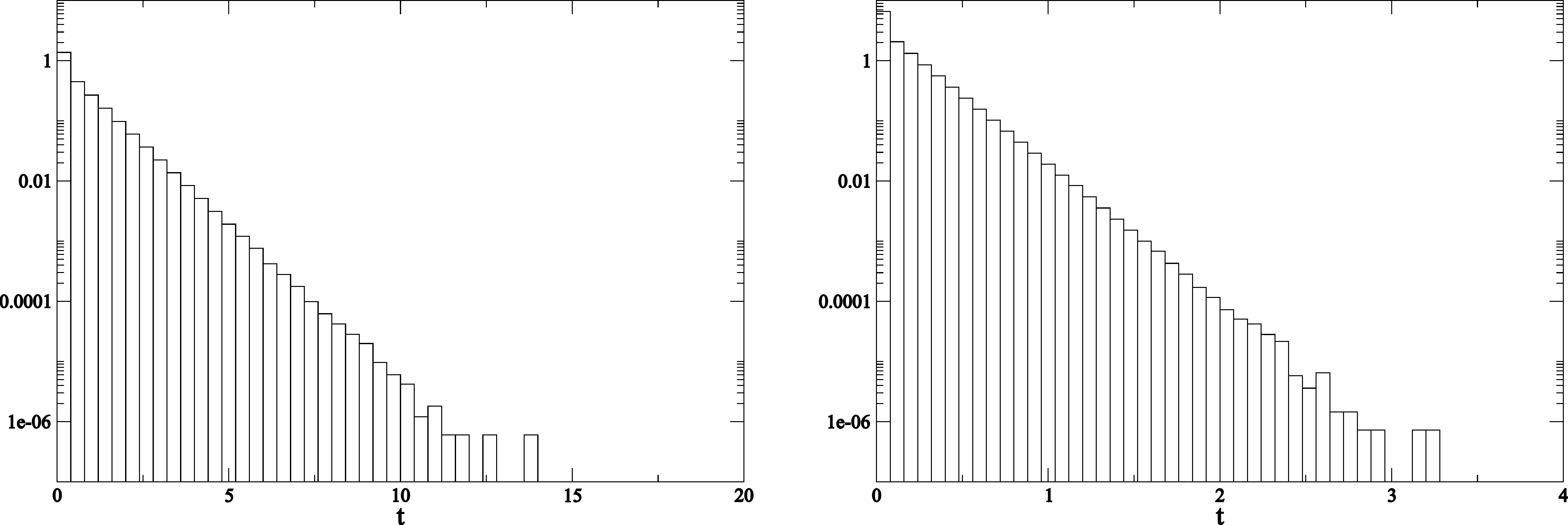}
\end{center}
\caption{a) Log-Linear plot of normalized histogram of sojourn times with cells in momentum space of width $\Delta p_x=10^{-4}$ and $N=2500$
and total simulation time $t_f=1000.0$.
b) Same as (a) but with $N=10\,000$ and $t_f=237.0$. Both histograms show clearly an exponential tail.}
\label{sojshort}
\end{figure}

\begin{figure}[ht]
\begin{center}
\includegraphics[width=16cm]{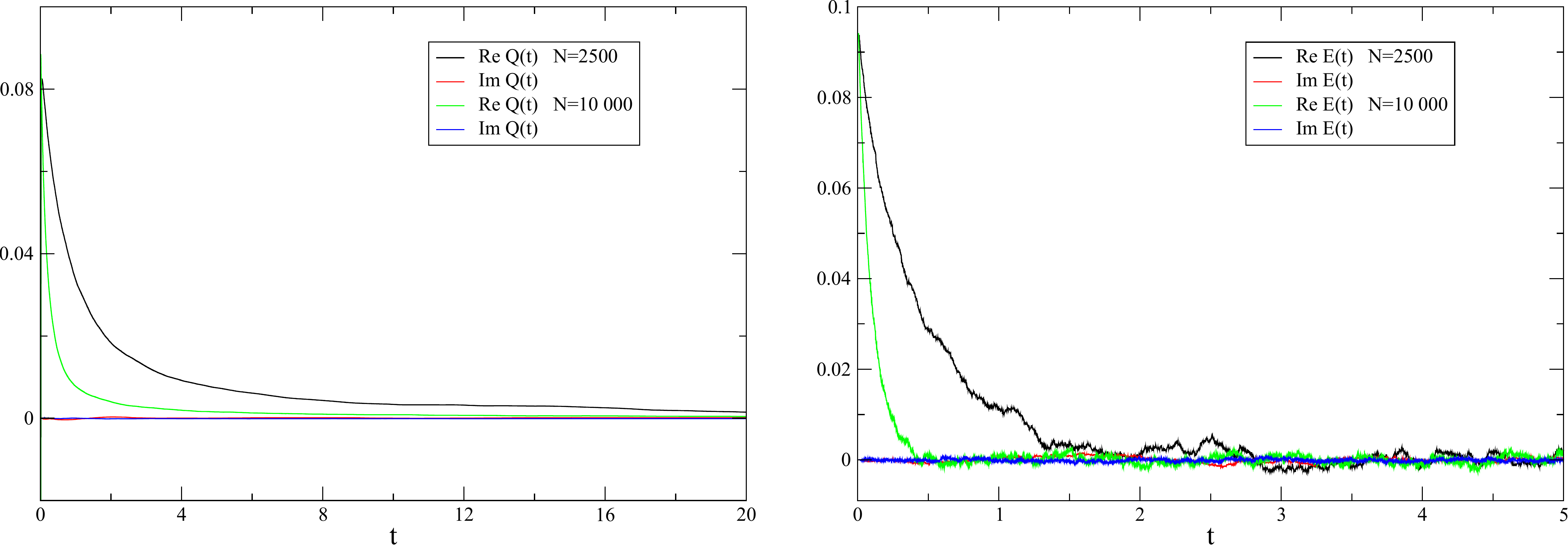}
\end{center}
\caption{(Color online) Left panel: Real and imaginary part of $Q(t)$ for a gas of hard-discs with $N=2500$ and $N=10\,000$ for the same initial
conditions as in Fig.~\ref{sigshort}. Right panel: Evolution of $E(t)$. In both cases we set $n=t$ in Eq.~(\ref{efen}).}
\label{ergomix}
\end{figure}

Sojourn times are obtained by considering a momentum cell size of $\Delta p_x=10^{-4}$.
The histograms (distribution functions) of sojourn times are shown in Fig.~\ref{sojshort} for $N=2500$ and $N=10\,000$.
An exponential tail is evident indicating that the system is ergodic (no power law in the tail).
Finally for the dynamical functional method, Fig.~\ref{ergomix} shows the real and imaginary parts of $E(n)$ and $Q(n)$.
Were we set $n=t$ for integral values of $t$.
The asymptotic values rapidly tend to zero in accordance with Eqs.~(\ref{ergodiccond2}) and~(\ref{mixcond2}) (up to some fluctuations),
indicating that the system is not only ergodic but also mixing.

\section{The Hamiltonian Mean-Field Model Revisited}
\label{hmfrev}

\begin{figure}[ht]
\begin{center}
\includegraphics[width=16cm]{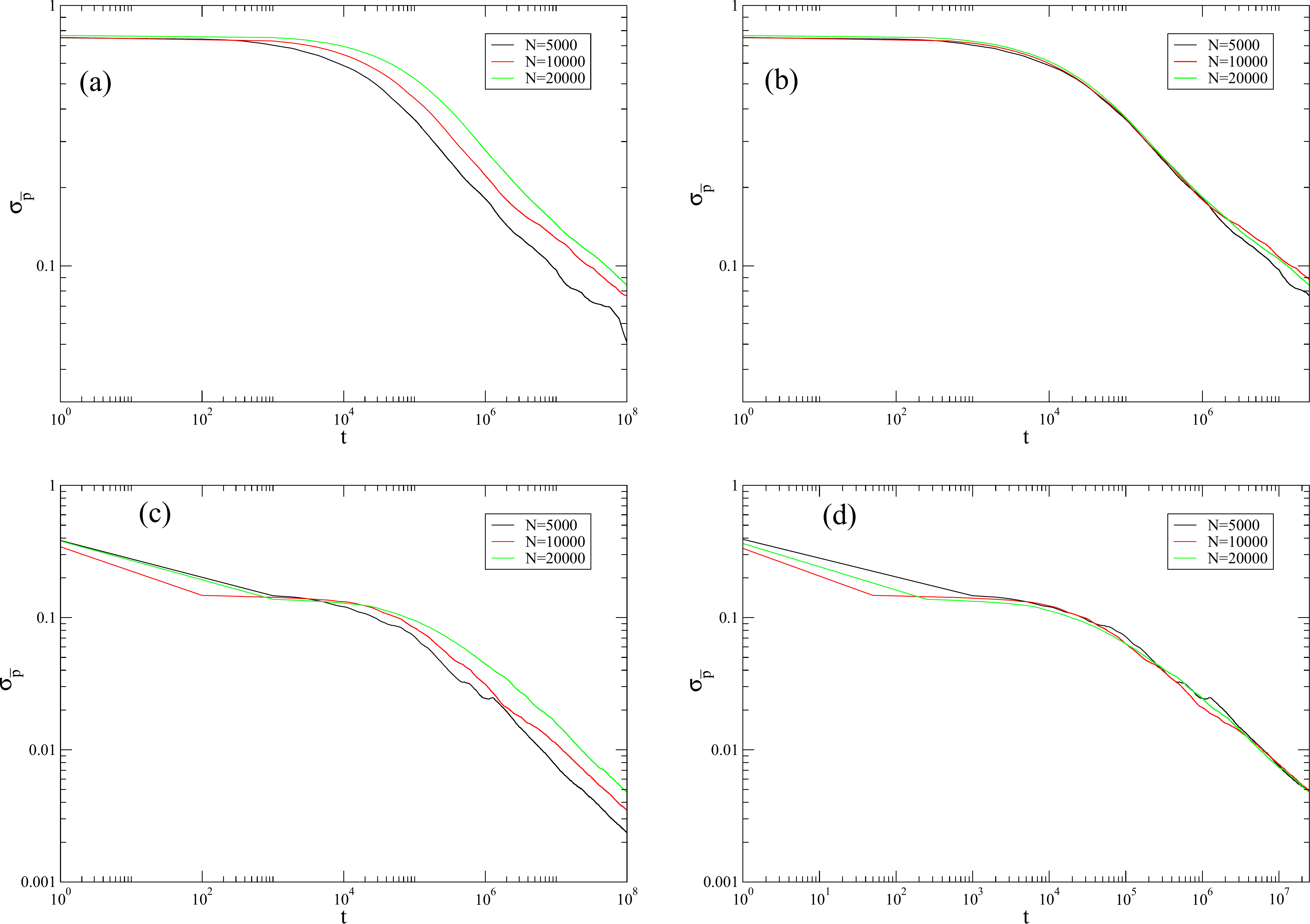}
\end{center}
\caption{Standard deviation $\sigma_{\overline{p}}$ in Eq.~(\ref{drt3}) for the momentum time averages for the HMF model for a few values of $N$ at thermodynamic equilibrium.
a) Total energy per particle $e=0.4$ (magnetized state). b) Same as (a) but the time rescaled as $t\rightarrow t\times5000/N$.
c) Total energy per particle $e=0.8$ (homogeneous state). d) Same as (c) but the time rescaled as $t\rightarrow t\times5000/N$.}
\label{fighmfsigmas}
\end{figure}

To better grasp how diverse is the behavior of systems with long-range interactions, we briefly discuss the ergodicity of the widely studied
HMF model. Its ergodic properties were studied in Ref.~\cite{prezol} using the sojourn times method
only for homogeneous non-equilibrium QSS. Here we discuss its ergodic properties for homogeneous and a 
non-homogeneous states at the thermodynamics equilibrium. We again restrict ourselves to analyze momentum variables
as position variables show a similar behavior, leaving a more complete analysis for the two-dimensional self-gravitating system in next section.
Figure~\ref{fighmfsigmas} shows the standard deviation $\sigma_{\overline{p}}$ in Eq.~(\ref{drt3}) for two different values of the total energy
per particle at thermodynamic equilibrium. For $e=0.4$ the inverse temperature is $\beta\approx3.02$ and the system is in a non-homogeneous state,
while for $e=0.8$ the state is homogeneous with $\beta\approx1.67$ and a non-vanishing mean-field force. The thermal average velocities are $\overline{v}\approx0.58$ and
$\overline{v}\approx0.77$, respectively. We first observe that the time required for the standard deviation $\sigma_{\overline{p}}$ to decrease by two orders of magnitude
is roughly six order of magnitude bigger than the crossing time $t_{cr}=2\pi/\overline{v}$, i.~e.\ the time
required for a particles with average thermal velocity to make a full turn on the circle.
Figures~\ref{fighmfsigmas}b and~\ref{fighmfsigmas}d show that $\sigma_{\overline{p}}(t)$ scales proportional to the number of particles $N$.
This is somehow expected for the non-homogeneous case as the collisional contribution to the kinetic equation is proportional to $1/N$, but
not for the homogeneous case as in this case the collisional integral is proportional to $1/N^2$. In fact it was show in Ref.~\cite{scaling}
that the statistical momenta of the one-particle distribution function, and therefore the distribution function itself, scale in time with
a factor $1/N$ for the non-homogeneous state and $1/N^2$ for the homogeneous case.
In the Vlasov limit the time for ergodicity diverges with $N$, at variance with what occurs for the short range system in the previous section.

\begin{figure}[ht]
\begin{center}
\includegraphics[width=16cm]{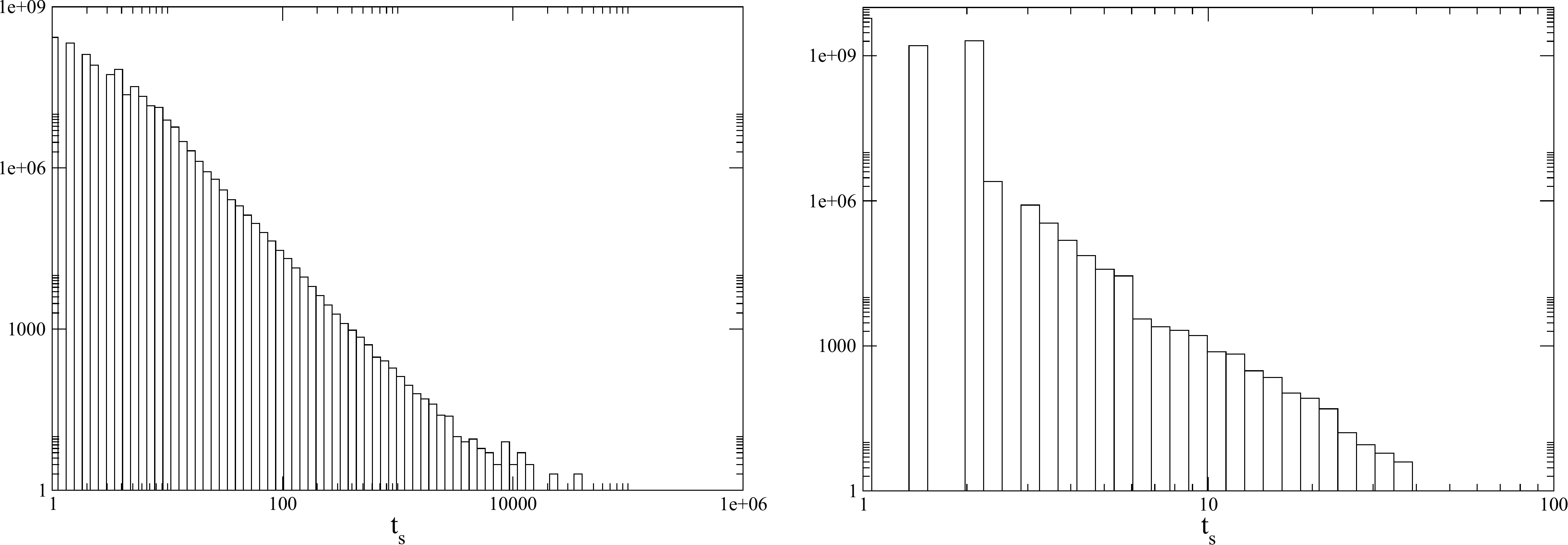}
\end{center}
\caption{Left Panel: Frequency histogram of sojourn times in momentum space for the HMF model for the same homogeneous state as
in Fig.~\ref{fighmfsigmas}a, with cell width $\Delta p=0.1$, total simulation time e$t_f=10^6$. Right Panel: histogram for the
non-homogeneous state in Fig.~\ref{fighmfsigmas}c, cell width $\Delta p=0.4$ and total simulation time $t_f=10^7$.
A truncated algebraic tail is visible in both cases.}
\label{hmfsoj}
\end{figure}

\begin{figure}[ht]
\begin{center}
\includegraphics[width=16cm]{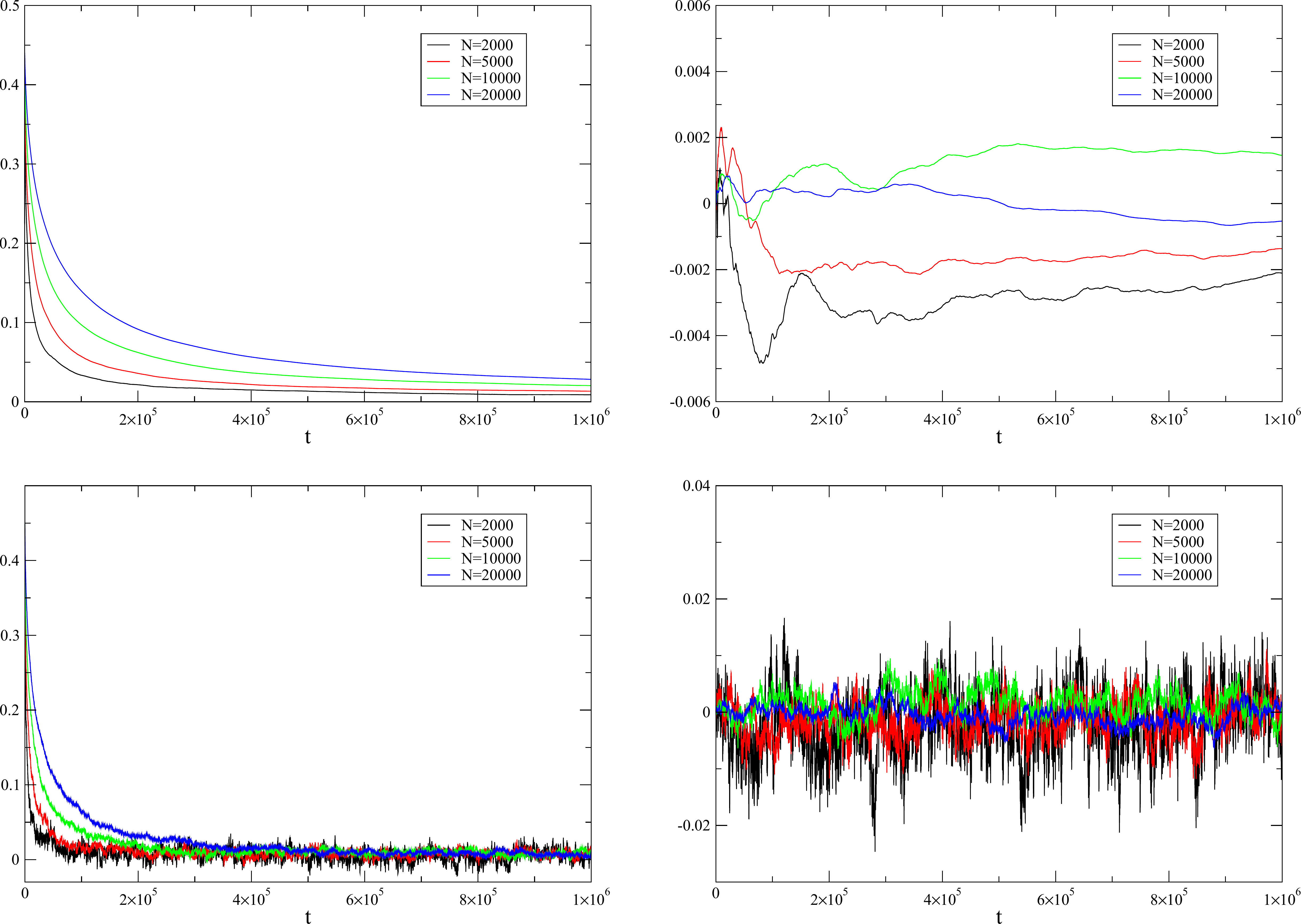}
\end{center}
\caption{Dynamical functional method for the HMF model in a homogeneous equilibrium state for the same state as in Fig.~\ref{fighmfsigmas}a. a) ${\rm Re}\left[Q(t)\right]$;
b) ${\rm Im}\left[Q(t)\right]$; c) ${\rm Re}\left[E(t)\right]$; d) ${\rm Im}\left[E(t)\right]$. In all cases we used $n=t$ with integer values of $t$.}
\label{hmfhomynf}
\end{figure}

\begin{figure}[ht]
\begin{center}
\includegraphics[width=16cm]{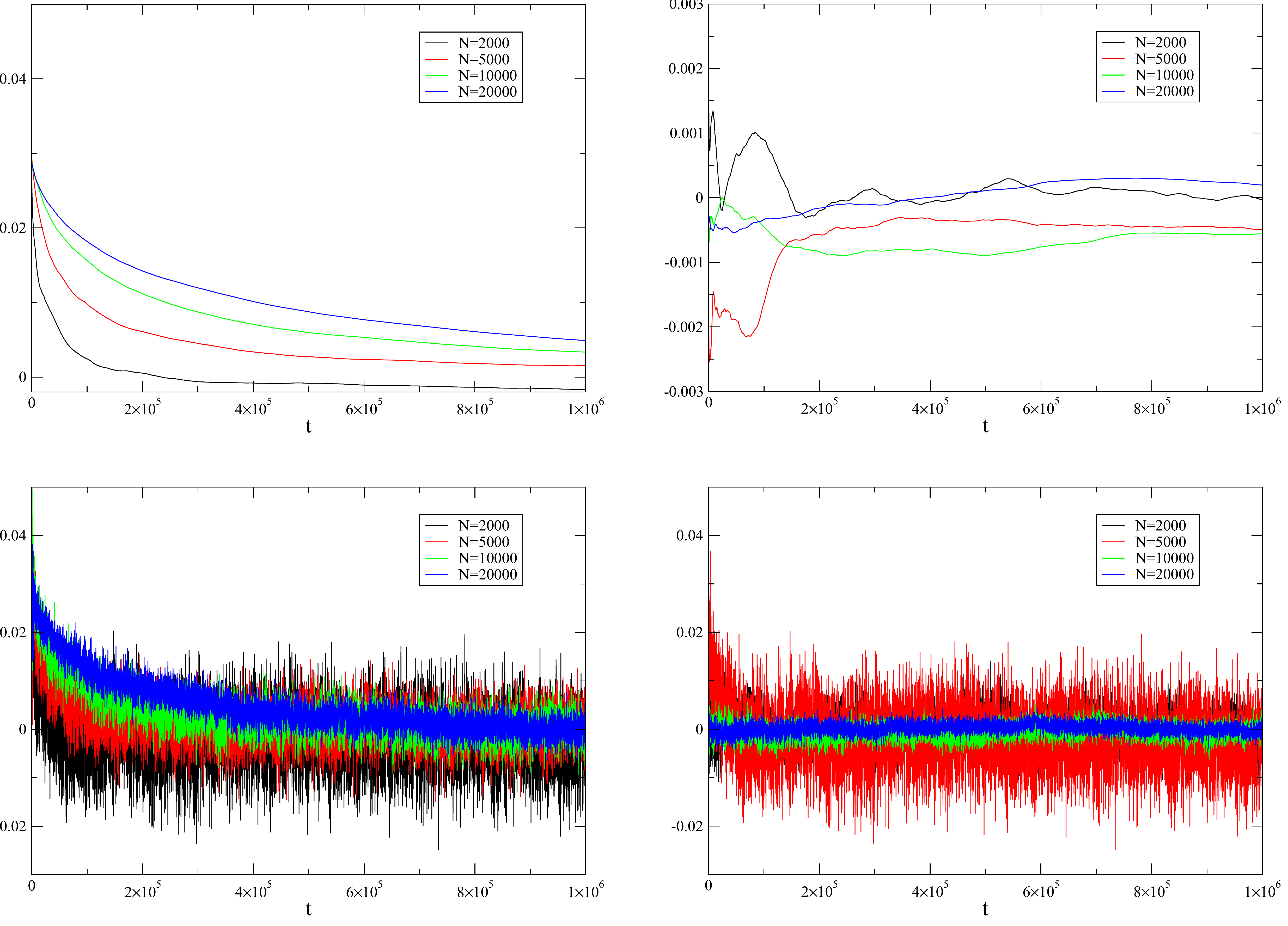}
\end{center}
\caption{Dynamical functional method for the HMF model in a non-homogeneous equilibrium state for the same state as in Fig.~\ref{fighmfsigmas}c. a) ${\rm Re}\left[Q(t)\right]$;
b) ${\rm Im}\left[Q(t)\right]$; c) ${\rm Re}\left[E(t)\right]$; d) ${\rm Im}\left[E(t)\right]$.}
\label{hmfhetdynf}
\end{figure}

\begin{figure}[ht]
\begin{center}
\includegraphics[width=16cm]{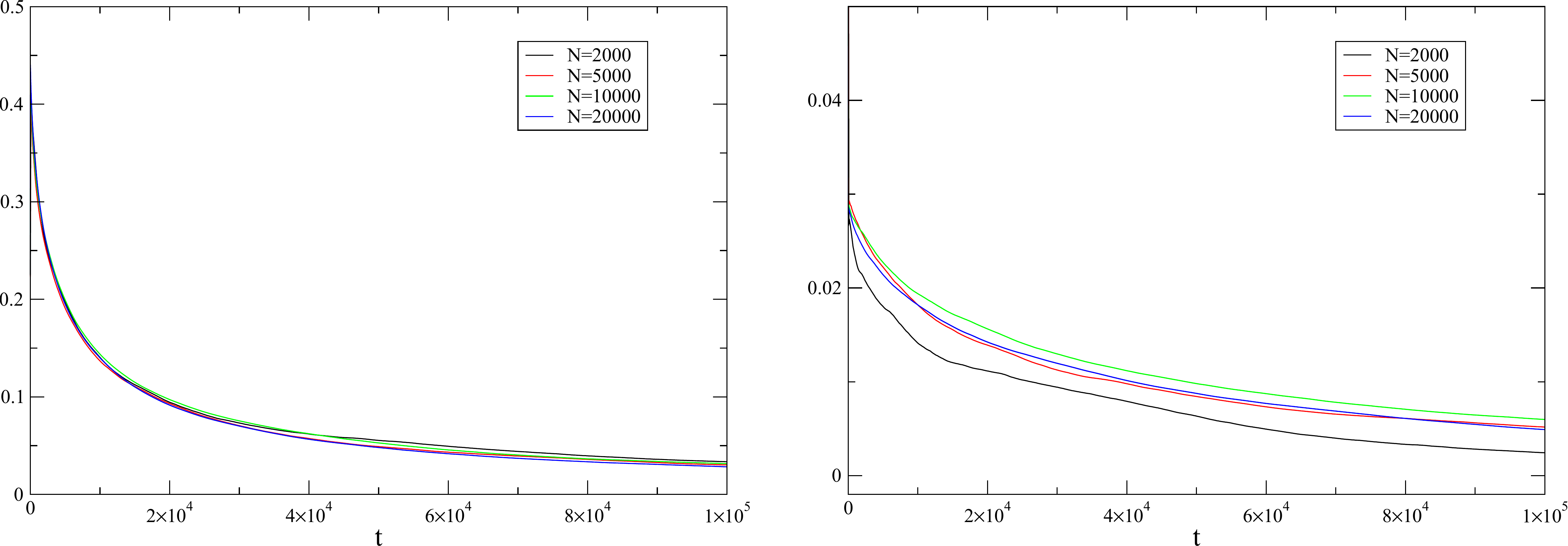}
\end{center}
\caption{Left and right panels are the same as in Figs.~\ref{hmfhomynf}a and~\ref{hmfhetdynf}a, respectively, but with time rescaled as
$t\rightarrow t\times N/5000$ in both cases. For the homogeneous case (Left panel) the data collapse is very good,while for the
inhomogeneous case (right panel), data collapse is only approximate.}
\label{hmfqncoll}
\end{figure}

Figure~\ref{hmfsoj} shows the distribution of sojourn times for the same initial states. A truncated algebraic tail is visible in the plot, in accordance
with Ref.~\cite{prezol}.
Results for the dynamical functional method are shown in Figs.~\ref{hmfhomynf} and~\ref{hmfhetdynf}, and we can conclude that the system is not only
ergodic but also mixing as long as $N$ is finite. To stress this point Fig.~\ref{hmfqncoll} shows that $Q(t)$ and $E(t)$ scale with $N$, and
similarly to the previous paragraph, the conditions for ergodicity and for mixing are not satisfied in the $N\rightarrow\infty$ limit.

\section{Two-dimensional gravity}
\label{tdg}

\begin{figure}[ht]
\begin{center}
\includegraphics[width=16cm]{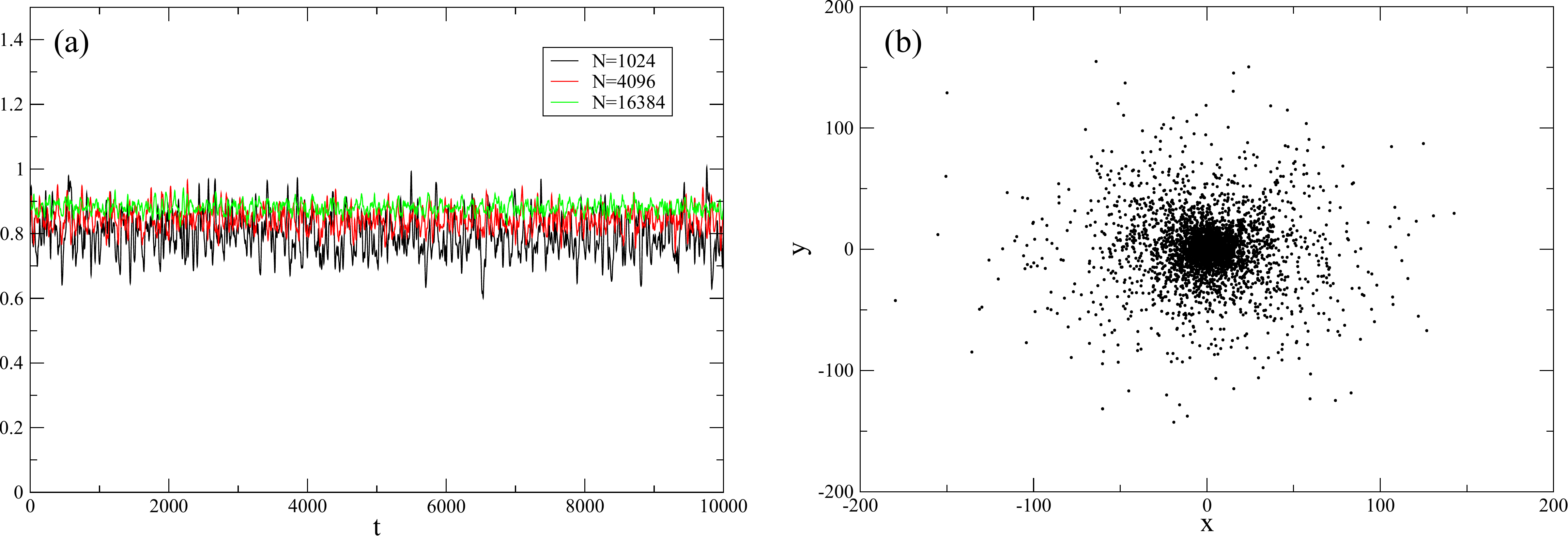}
\end{center}
\caption{a) Kurtosis for the momentum distribution as a function of time for the values of $N$ used below.
b) Positions of particles in the configuration used as initial conditions for simulations for $N=4096$.}
\label{momplot}
\end{figure}

For a self-gravitating $N$-body system with Hamiltonian in Eq.~(\ref{ham2dgrav}),
we consider an initial condition with particles distributed in space uniformly on a circular strip, with inner radius $R_1=40.0$
and outer radius $R_2=50.0$, and a uniform (waterbag) distribution for the momenta in the interval $-p_0<p<p_0$ with $p_0=0.5$.
The system then evolves for a total time $t_f=10\:000$ such that it settles in a QSS, which is then used as initial condition
for the different tests.
Figure~\ref{momplot} shows the Kurtosis of the momentum distribution, defined by ${\cal K}\equiv\langle p^4\rangle/\langle p^2\rangle^2$
(${\cal K}=3$ for a Gaussian distribution),
for the values of $N$ used in our simulations, showing that the statistical state of the system is stationary, up to fluctuations, as required.

\begin{figure}[ht]
\begin{center}
\includegraphics[width=16cm]{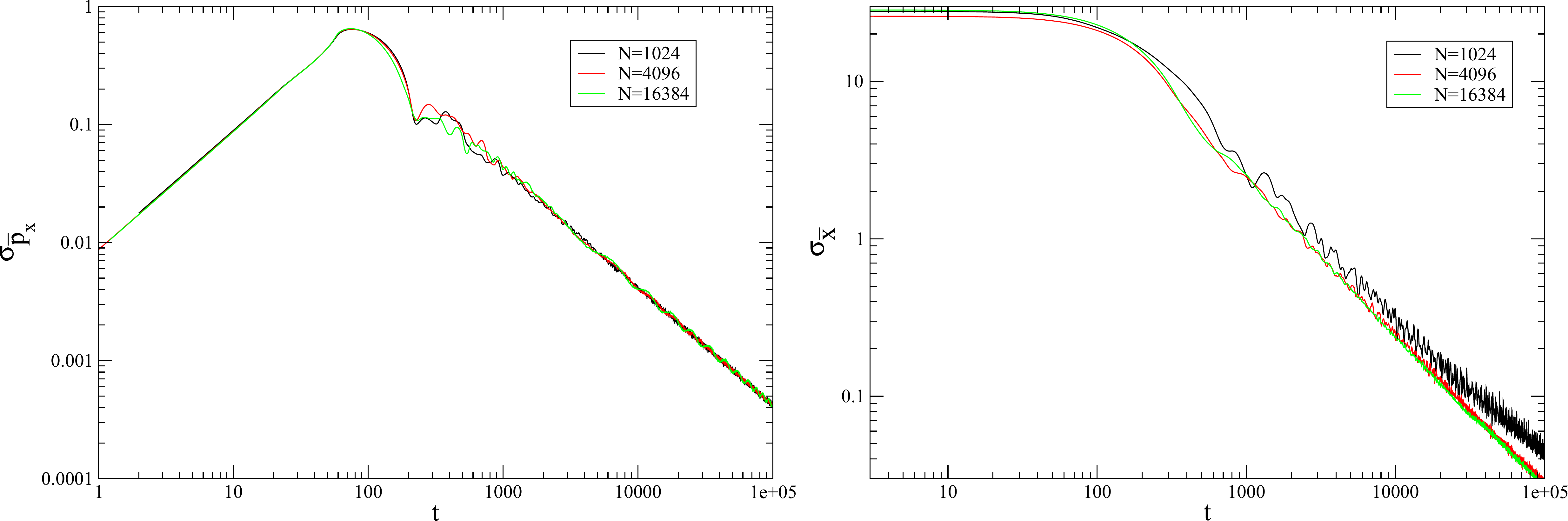}
\end{center}
\caption{(color online) a) Standard deviation in Eq.~(\ref{drt3}) for the variables $\overline{p_x}$ of each particle for the two-dimensional self-gravitating system,
integration time step $\Delta t=0.05$ and softening parameter $\epsilon=10^{-3}$. b) Standard deviation of $\overline{x}$.
The least-squares fit for the algebraic tail of $\sigma_{\overline{p}_x}(t)$ is given  by $48.4\times t^{-1.02}$ and
$2271\times t^{-0.99}$ for $\sigma_{\overline{x}}(t)$.}
\label{sigsum}
\end{figure}

The standard deviation of the time average of the $x$-components of momentum and position, as defined by Eqs.~(\ref{drt3})
and~(\ref{drt3b}), respectively, are shown in Fig.~\ref{sigsum}.
The average thermal velocity  is of order $\overline{v}\sim1$, and a crossing time can be defined by
the time a particle with the thermal velocity takes to cross the spatial extent $L_{qss}$ of
the core-halo QSS structure formed after the initial violent relaxation,
and shown in Fig.~\ref{momplot}b. By inspection, we obtain that $L_{qss}\approx200$. This characteristic length does not change significantly with $N$ and time.
The crossing time case is then given by $t_{cr}=L_{qss}/\overline{v}\approx 200$.
Comparing Fig.~\ref{sigsum} with Figs.~\ref{fighmfsigmas} and~\ref{sigshort}, we observe that the time for
$\sigma_{\overline{p}_x}$ and $\sigma_{\overline{x}}$ to decrease by two orders
of magnitude are roughly the same, and much smaller than for the HMF model in units of crossing time, but still much larger than for the short-range system.

The plots of $\sigma_{\overline{p}_x}(t)$ and $\sigma_{\overline{x}}(t)$ coincide after a small transient for all values of $N$.
The same behavior is observed for one-dimensional self-gravitating systems (not shown here).
This seems to imply that the time for ergodicity is the same for all $N$ and that two-dimensional self-gravitating systems are ergodic
even in the $N\rightarrow\infty$ limit. A few comments are in need on this point. The power law decay of the dispersion of $\overline{p}_x$ can be explained by
the fact that the final dispersion of the traveled distance in the $x$ direction
$\sqrt{\sum_k [x_k(t)-x_k(0)]^2/N}$ is roughly independent of $N$, and constant in time,
which implies that the exponent $\mu$ in Eq.~(\ref{drt3d}) vanishes after the violent relaxation. Equation~(\ref{drt1d})
then implies that $\sigma_{\overline{p}_x}$ is proportional to $t^{-1}$ as shown in Fig.~\ref{sigsum}.
This is at variance to what was obtained for the HMF model where diffusion never saturates~\cite{antoni2}.
There is no similar reasoning for the behavior of $\sigma_{\overline{x}}$ as already explained in Sec.~\ref{dirmeth}.
To explain the independence of $\sigma_{\overline{p}_x}(t)$ on $N$ we note that its time evolution is due to fluctuations of the force (for finite $N$)~\cite{chandra1,chandra2,kandrup}.
The statistics of the force fluctuations for a three-dimensional self-gravitating system were first obtained by Chandrasekhar~\cite{chandra2} and
for one and two-dimensional gravity by Chavanis~\cite{chavanisepj}. Chavanis showed that
the fluctuation of the force scale with the density of mass, which is proportional to $N$. Since he does not use a Kac factor in his expressions, we conclude
for the potential in Eq.~(\ref{ham2dgrav}) the fluctuations of the force are independent of $N$, and thus the same being true for $\sigma_{\overline{p}_x}(t)$.
In our case, if a Kac factor is not introduced, then the observation time required for the system be ergodic diverges with $\sqrt{N}$
(see discussion after Eq.~\ref{ham2dgrav}), i.~e.\ the system is strictly non-ergodic only in the $N\rightarrow\infty$ limit.
With a Kac factor the system is always ergodic and mixing, although a considerable  observation time is required.
It is also worth noticing that the scaling of $\sigma_{\overline{p}_x}$ with $N$ is not the same as for the dynamics of the one-particle distribution function,
which scales with $N$.

\begin{figure}[ht]
\begin{center}
\includegraphics[width=16cm]{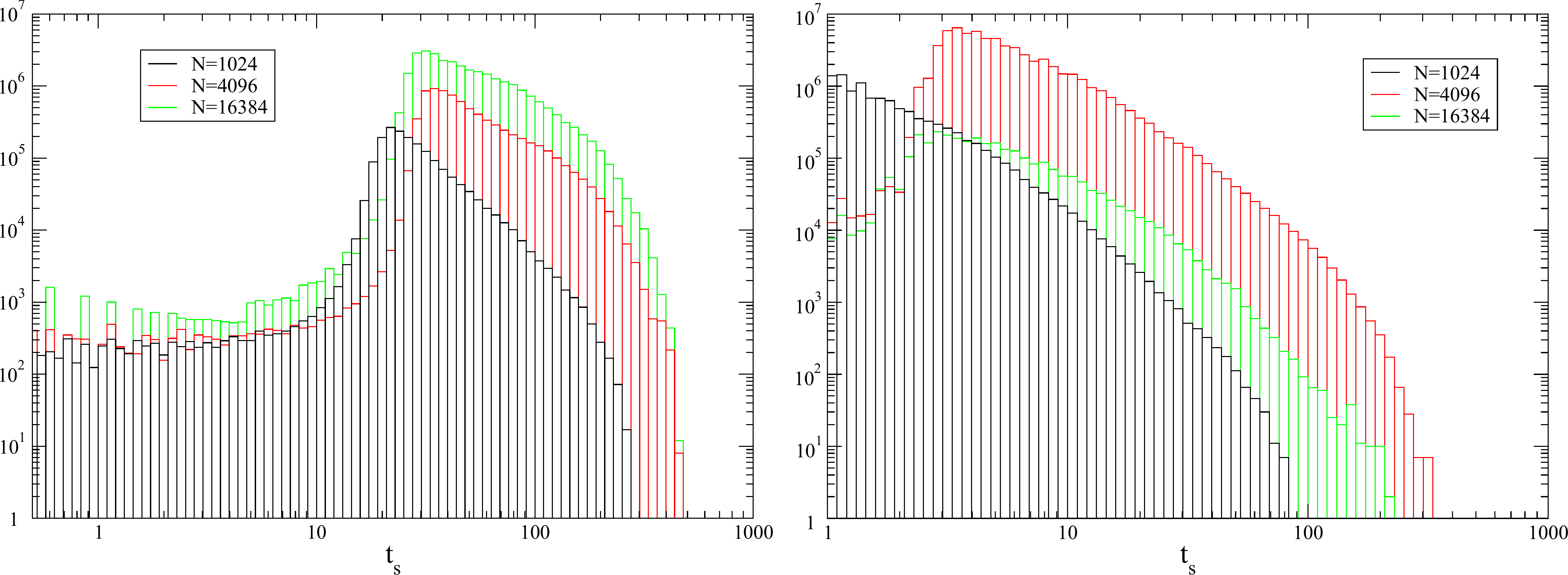}
\end{center}
\caption{a) Sojourn times for cells in momentum $x$ component of momentum space with width $\Delta p_x=0.2$ for some different values of $N$
and total simulation time $t_f=10^4$.
b) Same as (a) but with cells in the $x$ component of the position with width $\Delta x=0.2$ and $t_f=10^5$.}
\label{S1}
\end{figure}

Sojourn times are again obtained by diving the momentum space into equal size cells of fixed width, associated with discrete states.
First we consider cells in momentum and position $x$ components, with width $\Delta x=0.2$ and $\Delta p_x=0.2$, respectively.
Figure~\ref{S1} shows the frequency histograms for sojourn times $t_s$, which is truncated at quite a small value.
This truncation is expected to occur for finite systems as discussed in Ref.~\cite{prezol}. On the other hand
the truncation values does not vary significantly with $N$, at variance with what was observed for the HMF model.
An algebraic tail is roughly visible for the intermediate values of $t_s$, and deviates from it near the truncation of the distribution, were the statistics is less significant.

\begin{figure}[ht]
\begin{center}
\includegraphics[width=16cm]{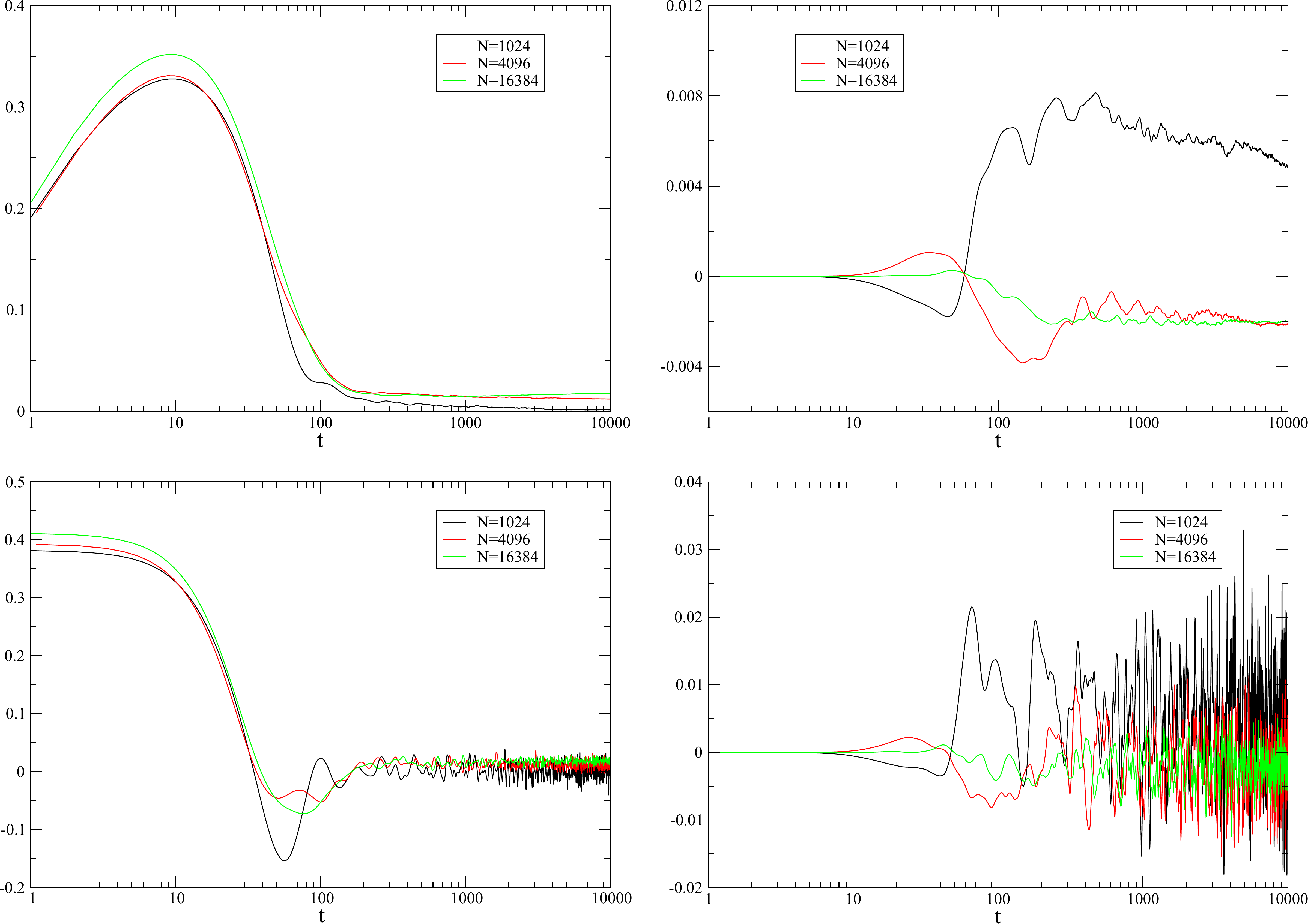}
\end{center}
\caption{a) Real and b) imaginary part of $Q(n)$ in Eq.~(\ref{ergodiccond2}) for the same simulation and number of particles $N$ as in Fig.~\ref{S1}
for the $x$ component of the momentum as stochastic variable.
c) Real and d) imaginary part of $E(n)$ in Eq.~(\ref{mixcond2}).}
\label{dynfunc1}
\end{figure}

\begin{figure}[ht]
\begin{center}
\includegraphics[width=16cm]{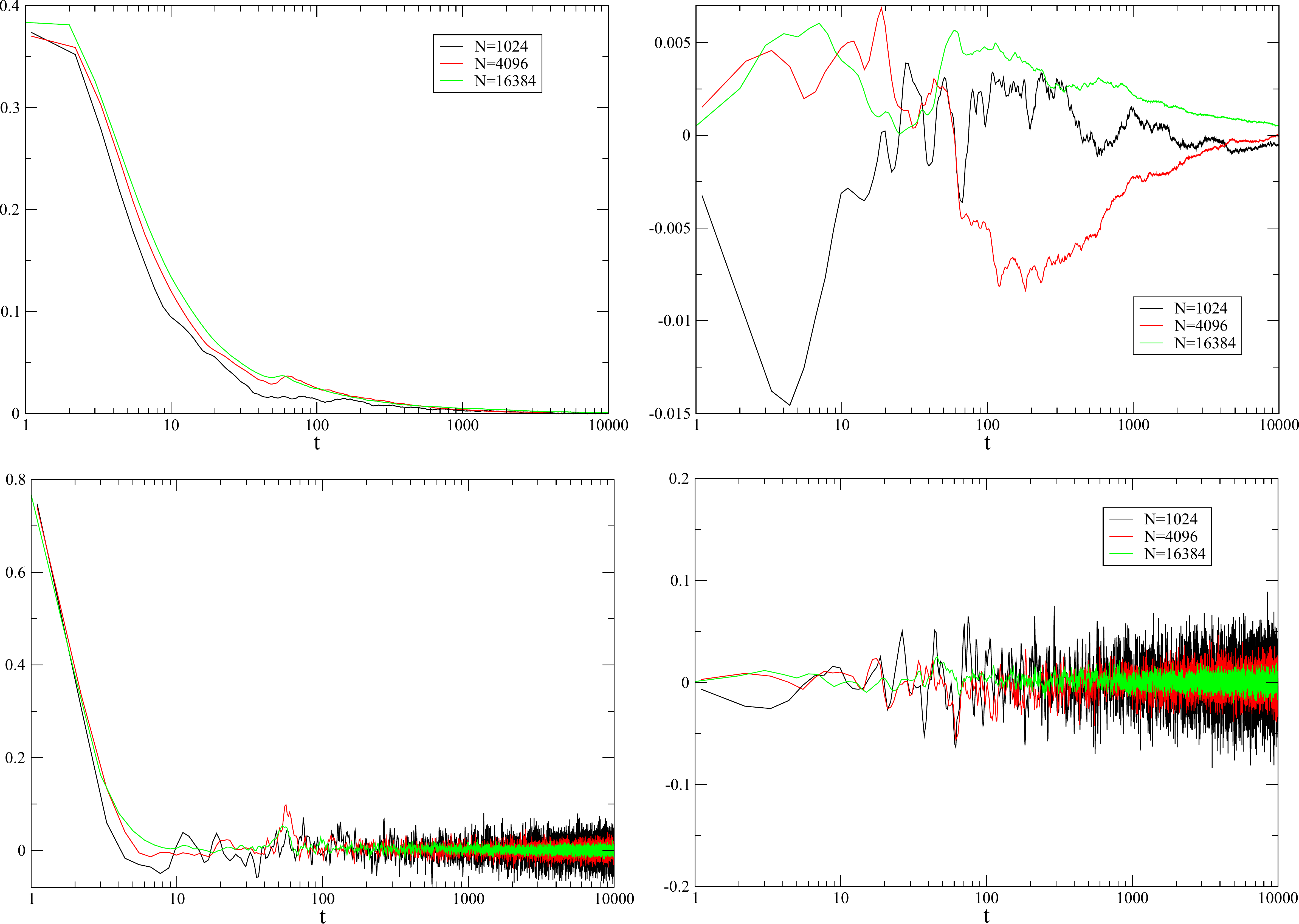}
\end{center}
\caption{Same as Fig.~\ref{dynfunc1} but For the $x$ component of position as stochastic variable.}
\label{dynfunc2}
\end{figure}

Results for the dynamical functional method are given in Figures~\ref{dynfunc1} and~\ref{dynfunc2}, showing
$Q(n)$ and $E(n)$, with $n=t$ for integer values of $t$. In all cases
Eqs.~(\ref{ergodiccond2}) and~(\ref{mixcond2}) are satisfied only after a very long time (compare with Fig.~\ref{ergomix})
which is roughly the same for all values of $N$, in agreement with our results above.

\section{Concluding remarks}
\label{cr}

We discussed the ergodic and mixing properties of a two-dimensional self-gravitating system using the methods
described in Section~\ref{te}. The potential of the system is regularized by the Kac factor in order to have a well
defined thermodynamic (or Vlasov) limit. In order to better understand our results we also applied the same tests to
a gas of two-dimensional hard-discs, and to the HMF model. We showed that all these models are ergodic
and mixing, although for the HMF model this is true only if $N$ is finite,
as the observation time required for the system to be ergodic (and mixing) diverges with $N$. Nevertheless, for two-dimensional gravity,
this time scale seems to be independent on $N$. This is related to the confining nature of the interaction potential and the
scaling of fluctuations of the force around its mean-field value on $N$. If a Kac factor is not introduced in the potential, then the time for ergodicity diverges
with increasing $N$. For the hard-disc gas, the time for ergodicity goes to zero in the thermodynamic limit, at variance with both long-range
interacting systems studied here.

The direct and dynamical functional methods proved to be
relevant tools for testing ergodicity for the self-gravitating system, while the sojourn times approach gave more ambiguous results, as the
tail of their distribution is too short in the present case to asses if there is an algebraic part.

We note that up to the authors knowledge, the only previous study where such characteristic time was estimated for long-range
interacting systems were devoted to the Hamiltonian mean field model, where a $N$ dependence was reported~\cite{nosepl,prezol}.
This result is reobtained here for an equilibrium initial condition.

We also point out that a self-gravitating system with $N\rightarrow\infty$ is weakly non-ergodic since in the one-particle phase space the constant energy hypersurface is always connected,
and all regions are necessarily accessible by the dynamics. In this way, a mean-field system can strongly non-ergodic if the corresponding energy
hypersurface can be composed, at least for some energy values, by disjoint sets.

\section{Acknowledgments}

TMRF was partially financed by CNPq and CHS was financed by CAPES (Brazilian Government agencies).

\end{document}